\documentclass[12pt,a4paper]{article}
      \usepackage{amssymb}
      \def\pa{\partial}
      
      \def\k{\kern-.1em\mathbin{,}\kern-.1em}
      \def\hk{\kern.12em\raise-1em\hbox{$\hat{\raise1em\hbox{,}}$}\kern.12em}
      
      \newcommand{\initiate}{\setcounter{equation}{0}}

      \def\be{\begin{equation}} \def\bea{\begin{eqnarray}}
      \def\ee{\end{equation}}\def\eea{\end{eqnarray}}
      \def\ncr{\nonumber\\ }

      \def\slash{{\rlap /}}
      
      \def\be{\begin{equation}} \def\bea{\begin{eqnarray}}
      \def\ee{\end{equation}}\def\eea{\end{eqnarray}}
      \def\ncr{\nonumber\\ }
      \def\STr{{\rm STr}\,}
      \def\Sdet{{\rm Sdet}\,}\def\Tr{{\rm Tr}\,}

      \def\MBVR{Maja Buri\'c\footnote{E-mail: majab@ff.bg.ac.yu}
      \ and
      Voja Radovanovi\'c\footnote{E-mail: rvoja@ff.bg.ac.yu}\\
      {\it Faculty of Physics, P.O. Box 368, 11001 Belgrade,
      Serbia and Montenegro}}

      \def\endtitle{\par\end{quotation}\vskip3.5in minus2.3in\newpage}

      \def\a{\alpha}
      \def\b{\beta}
      
      \def\d{\delta}
      \def\e{\epsilon}                
      \def\g{\gamma}

      \def\k{\kappa}
      \def\l{\lambda}
      \def\m{\mu}
      \def\n{\nu}
      
      \def\r{\rho}                    
      \def\s{\sigma}                  

      \def\D{\Delta}
      
      \def\G{\Gamma}

      \def\pa{\partial}

      \def\ca{{\cal A}}
      \def\cb{{\cal B}}
      \def\cc{{\cal C}}

      \def\ci{{\cal I}}

      \def\cm{{\cal M}}

\begin{document}

      \title{Non-renormalizability of the noncommutative $SU(2)$ gauge theory}

      \author{\MBVR}

      \date{}

      \maketitle

      \abstract{We analyze the divergent part of the one-loop effective action
      for the noncommutative $SU(2)$
       gauge theory coupled to the fermions in the fundamental representation.
       We show that the divergencies in the 2-point and the 3-point functions
       in the $\theta$-linear order can be renormalized, while the divergence in the 4-point
       fermionic function cannot.}

      \vfill \noindent  \eject

      \parskip 4pt plus2pt minus2pt

      \initiate \section{Introduction}

      Although discussed for quite some time, the question of
      renormalizability of field theories on noncommutative ${\bf R}^4$
      has not been settled in a satisfactory way yet. Noncommutativity
      of the coordinates, i.e., a relation of the type
      \be\left[ \hat x^\m , \hat x^\n \right] = i\theta ^{\m\n},
      \label{canonical}\ee
      puts the
      lower bound on the coordinate measurements, so one would expect
      that it also implies a natural ultraviolet cut-off and acts as a
      regulator. However, this idea has not been successfully
      implemented in models.

      Usually one represents the noncommutative field $\hat \Phi (\hat
      x)$ by a function $\hat\Phi (x)$ on ${\bf R}^4$ and encodes
      the
      noncommutativity in the multiplication rule ($\star$-product).
      For example, for  constant
      $\theta ^{\m\n}$, the field multiplication is given by the
      Moyal-Weyl product: \be \label{moyal} \phi (x)\star \chi (x) =
      e^{\frac{i}{2}\,\theta^{\m\n}{\pa\over \pa x^\m}{\pa\over \pa
      y^\n}}\phi (x)\chi (y)|_{y\to x}\ .\ee This product is nonlocal,
      and so is the field theory defined by it.

      The most extensive study of  renormalizability of  noncommutative
      (NC) field theories was done for the scalar field theory
      \cite{phi4}. One quantizes perturbatively: the diagrams in the
      perturbation theory split into planar and  nonplanar. The planar
      diagrams reproduce the behavior of the underlying commutative
      theory; on the other hand, UV divergencies in the nonplanar graphs
      get regulated by the effective cut-off $(\theta p)^{-1}$. But for
      the exceptional values of the momenta (when the momentum flow into
      the loop is zero), these contributions become infinite. The
      reappearance of the UV divergencies in the infrared sector is
      related to nonlocality of the theory. Renormalizability of
      noncommutative $\Phi ^4$ has been analyzed in the recent papers
      \cite{gp,becchi,gw} from the point of view of the
      Wilson-Polchinski RG equation. The renormalization procedure was
      defined in some special cases; it was shown that this procedure is
      different from the usual planar renormalization.

      Noncommutative gauge theories, in particular $U(1)$ and $U(N)$,
      have been studied on
      the similar lines, too. The UV/IR mixing  appears in the Feynman
      graphs in the same way as for the scalar field theory
      \cite{U1,UN}, so the question of renormalizability has the same
      status. However, for the gauge theories there is another
      representation. As shown by Seiberg and Witten \cite{sw},
      noncommutative and  commutative gauge theories are equivalent.
      This equivalence  is realized by a mapping relating the
      representation (gauge fields, matter fields) of NC gauge symmetry
      to the fields carrying the representation of its commutative
      counterpart. The SW map  is given as a series in powers of
      $\theta^{\m\n} $. Classical action is also expanded in $\theta$:
      in the zero-th order it reduces to the action of ordinary gauge
      theory; additional terms can be treated as couplings. Nonlocality
      shows up in the infinite number of interactions.
      This realization gives another framework to address
      the issue of renormalizability: it in particular makes sense in
      the limit of small noncommutativity, $\theta \to 0$. One of the
      features of '$\theta$-expanded' approach is that arbitrary gauge
      groups and tensor products can be represented, so  noncommutative
      generalizations of the standard model have been constructed
      \cite{smodel}.

      In this paper, we study the renormalizability of NC $SU(2)$ in the
      $\theta$-expanded approach. Renormalizability of the
      $\theta$-expanded NC gauge theories has been addressed in papers
      \cite{u1,w,susy,mv} for the gauge group $U(1)$ with or without fermionic
      matter and for its
      supersymmetric extension. Here we discuss the $SU(2)$
       theory coupled to
      fermions in the fundamental representation. Although the $SU(2)$ gauge
      theory  technically differs from the abelian $U(1)$ in many
      details, the conclusion concerning renormalizability is the same.
      If fermions are massive, theory is not renormalizable. For
      massless fermions the theory is  "almost renormalizable", meaning
      that there is only one divergent term in the effective action
      which cannot be absorbed by the SW field redefinition scheme. The
      method which we used to calculate the divergent contributions  is
      the background field method. As it was thoroughly explained in
      \cite{mv}, we skip the technical points here, referring also to
      the standard literature \cite{peskin}. The calculations for
      $SU(2)$ are quite involved in comparison with $U(1)$, so in
      this paper we discuss only the $\theta$-linear order and find the
      divergent parts of the 2-point, 3-point and fermionic 4-point
      functions. Previously, the results for $U(1)$ in the
      $\theta$-linear order were given in \cite{u1,w}; for the 2-point
      functions they were extended in \cite{mv} to the $\theta
      ^2$-order.

      The plan of the paper is the following. In the second section we
      describe the classical action for noncommutative $SU(2)$. In the
      third section all necessary steps for the perturbative
      quantization are done. The results for the divergencies of the
      2-point, 3-point and fermionic 4-point functions are given in the sections
      4 and 5. The results which are obtained and some further issues
      concerning renormalizability are discussed in the last section.

      \section{The model}

      The general construction of gauge theories on noncommutative
      space and their relation to the SW map were introduced in
      \cite{mad,mssw,jmssw}; we will repeat only a few relevant
      steps.

      The noncommutative space is an algebra generated by a set of
      noncommuting coordinates $\hat x^\m$; in general they obey
      relations, $\mathcal{R}(\hat x) = 0$; for example
      (\ref{canonical}). Physical fields $\hat\psi (\hat x)$ are
      functions of the coordinates. We want to describe the
       gauge theory: let
      $\a (x) = \a ^a(x)T^a$ be a (commutative) gauge parameter
      and
      $T^a$ -- generators
      of a Lie group. The field $\hat\psi (\hat x)$ transforms
      covariantly under the infinitesimal gauge transformation $\hat
      \Lambda_\a (\hat x)$ if $\, \hat \delta _{\a}\hat\psi (\hat x) =
      i\hat\Lambda_\a (\hat x)\hat \psi (\hat x)\ .$
      The
      gauge degrees of freedom are inner, so the coordinates $\hat x^\m$
      are invariant under gauge transformations: $\hat\delta _{\a}\hat
      x^\m = 0$. However, one can define the 'covariant coordinates'
      $\hat X^\m$ introducing the gauge potentials $\hat A_\m(\hat x)$:
      $\hat X^\m = \hat x ^\m + \theta^{\m\n}\hat A_\n(\hat x)$. They
      have the transformation property $\hat\delta _{\a}\hat X^\m =
      i\left[ \hat\Lambda_\a , \hat X^\m\right]$, when the vector
      potential transforms appropriately. In order that the infinitesimal gauge
      transformations close, one impose \be\hat
      \delta_{\a}\hat\delta_{\b}-\hat\delta_{\b}\hat\delta_{\a} =
      \hat\delta_{\a\times \b} ,\label{SWequ}\ee where $\a\times
      \b=-i[\a,\b]$ denotes the composition of two transformations.

      In principle, for a given set of relations $\mathcal{R}=0$, the
      noncommutative coordinates $\hat x^\m$ can be represented by the
      coordinates $x^\m$ on a commutative manifold, and the
      multiplication by some $\star$-product. As already mentioned, in
      the case of the constant noncommutativity (\ref{canonical}) (which
      we here consider), the $\star$-product is the Moyal product
      (\ref{moyal}); it is possible to construct  $\star$-products for
      the other cases, too. Expanding the gauge parameter and physical
      fields  in $\theta$, the condition (\ref{SWequ}) becomes an
      equation (Seiberg-Witten) for the coefficients in the expansion. A
      solution of the SW equation is
      \bea
      \hat\Lambda (x) &=&\alpha (x) +\frac 14 \, \theta
      ^{\m\n}\left\{ \pa _\m\alpha (x) ,A_\n(x) \right\}+\dots\ncr
      \hat A_\r(x) &=& A_\r(x) -\frac 14 \,\theta ^{\m\n}\left\{ A_\m(x), \pa
      _\n A_\r(x)
      +F_{\n \r}(x)\right\} +\dots\label{expansion}\\
      \hat\psi (x) &=& \psi (x)- \frac 12 \, \theta ^{\m\n}
       A_\m(x)\pa _\n\psi (x) +\frac
      {i}{4}\, \theta ^{\m\n}A_\m(x)A_\n(x)\psi (x)+\dots \ ;\nonumber\eea
     the terms of  the second order in $\theta$ are given in \cite{jmssw}. In
     the last formula
      $\alpha (x)$, $A_\m(x)$ and $\psi (x)$ denote the commutative gauge
      parameter, the vector potential and the fermionic field:
      \be
      \label{covar} A_\m = A_\m^aT^a\, ,\ \ D_\m \psi =\pa
      _\m\psi-iA_\m\psi \ ,\ee
      \be F_{\m\n} =
      \pa _\m A_\n -\pa _\n A_\m -i\left[ A_\m,A_\n\right]\ \ee
      which correspond to the noncommutative quantities.
      As  terms linear in $\theta$ in
      (\ref{expansion}) contain
      anticommutators of $T^a$, it is clear that NC
      gauge fields $\hat\Lambda (x)$, $ \hat A_\m(x) $ do not close into
      the Lie algebra; they take values in the enveloping algebra of the
      gauge group.

      The other property of the solution (\ref{expansion}) is that it is
      not unique. It can be shown \cite{u1,a} that, if $A_\mu^{(n)}$,
      $\psi ^{(n)}$ is a solution of the SW equation (\ref{SWequ}) up to
      the order n in $\theta$, then by adding  any  gauge covariant
      expression (of appropriate dimension) with  exactly $n$ factors of
      $\theta$ to $A_\mu^{(n)}$, $\psi ^{(n)}$ one obtains another
      solution. This is similar to the relation between the solutions of
      inhomogeneous and the corresponding homogeneous linear equation;
      in a way, it expresses the nonlocality of the theory. As pointed
      out in \cite{u1}, this nonuniqueness can be used to subtract
      the
      divergent terms in the effective action and to regularize the
      theory -- one can think of such a procedure as a sort of
      'dressing'
      of the 'bare' fields, that is, the choice of the physical ones.

      We now proceed to the action. For NC $SU(2)$ Yang-Mills
      theory the classical action is given with \be \label{Snc}S=\int d^4x\,\hat{\bar
      \psi}\star (i\g^\m\hat D_\m-m)\hat\psi -\frac 14 \int d^4x\,
      \Tr(\hat F_{\m\n}\star\hat F^{\m\n})\ ,\ee
      where the noncommutative
      field strength  $\hat F_{\m\n}$  is \be \hat
      F_{\m\n}=\pa_\m\hat A_\n-\pa_\n\hat A_\m-i(\hat A_\m\star\hat
      A_\n-\hat A_\n\star\hat A_\m)\ ,\ee and the covariant derivative
      \be \hat D_\m\hat\psi = \pa_\m\hat\psi -i\hat A_\m\star \hat \psi\
      .\ee The commutative counterpart $\psi (x)$ of the matter field
      $\hat\psi (x)$ is in the fundamental representation of $SU(2)$,
      $T^a = \frac{\sigma _a}{2}$.
       Inserting the expansion (\ref{expansion}) into  (\ref{Snc}),
       we get the action in the$\theta$-linear order  \cite{jmssw}:
      \be S =S_0 +S _{1,A}+S_{1,\psi}\ ,\ee
      \bea\label{l0} S _{0 }\ &=&\int
      d^4x\,\left(\bar\psi \big( i\g ^\m D_\m-m\big)\psi -{1\over
      4}F^{\m\n a}F_{\m\n}^a\right)\ ,\\
      S_{1,A} &=& 0 \ ,\ncr
      \label{l1psi}S_{1,\psi}
      &=&\frac 12\,\theta ^{\r\s}\int d^4x\,\left( -i\bar\psi\g ^\m
      F_{\m\r}D_\s\psi +\frac 12\, \bar\psi F_{\r\s}(-i\g ^\m
      D_\m+m)\psi \right)
      \\ &=&
      \int
      d^4x\left( -\frac 18\,\theta^{\r\s}\Delta^{\m\n\a}_{\r\s\b}\bar
      \psi F_{\m\n}\g^\b (i\pa_\a+A_\a)\psi
      +\frac{m}{4}\,\theta^{\m\n}\bar\psi F_{\m\n}\psi\right)\ .
      \nonumber \eea In order to simplify the notation we introduced the
      symbol  $\Delta^{\a\b\g}_{\s\r\m}$  defined as \be \Delta
      ^{\a\b\g}_{\s\r\m} =\d ^\a_\s\d^\b_\r\d^\g_\m -\d
      ^\a_\r\d^\b_\s\d^\g_\m +({\rm cyclic\ }\a \b \g  )=-\epsilon
      ^{\a\b\g\l}\epsilon _{\s\r\m\l}\ .\ee
      The bosonic $\theta$-linear term $ S_{1,A}$
      vanishes (unlike in the $U(1)$ case) because it is proportional to
      the symmetric coefficients $d^{abc}=\Tr(\{T^a,T^b\}T^c)$, and for
      $SU(2)$ these coefficients are zero for all irreducible
      representations.

      In the functional integration we will treat ($A_\m$, $\psi$)
      as a multiplet so we want both fields to be real (or to be complex).
      Therefore we write the Dirac spinor $\psi$ in terms of the
      Majorana spinors $\psi _{1,2}$. For the charge-conjugated spinor
      $\psi
      ^C=C\bar\psi^T$  Majorana spinors are given by $\psi _{1,2}=\frac
      12(\psi\pm\psi ^C)$; vice versa,  $\psi = \psi _1 +i\psi _2\,$
      \footnote{We
      encounter some of the identities which  Majorana spinors satisfy:
      $\bar\phi\chi =\bar\chi\phi$ ; $\bar\phi\g _\m\chi =-\bar\chi\g
      _\m\phi$; $\bar\phi\s _{\m\n}\chi =-\bar\chi\s_{\m\n}\phi$;
      $\bar\phi\g _5\chi =\bar\chi\g _5\phi$; $\bar\phi\g _\m\g _5\chi
      =\bar\chi\g _\m\g _5 \phi$. }. To express the action in terms of
      Majorana spinors, one has to use explicitly  the form of
      Pauli matrices (i.e., the representation of the group
      generators). As $\sigma _2$ is antisymmetric and $\sigma _1$,
      $\s_3$ are symmetric matrices,
      the gauge field $A_\mu^2$ couples to  Majorana spinors
      differently from  $A_\mu^1$, $A_\mu^3$.
      The action
        reads:
      \bea S _0\label{act1} &=& \int d^4x\, \Big(\bar\psi _1 \big(
      i\slash \pa -m +\slash A^2\frac{\s_2}{2}\big)\psi _1 + \bar\psi
      _2\big( i\slash \pa  - m+\slash A^2\frac{\s_2}{2} \big)\psi _2\ncr
      &+& i\bar\psi _1(\slash A^1\frac{\s_1}{2}+\slash
      A^3\frac{\s_3}{2})\psi _2-i\bar\psi _2(\slash
      A^1\frac{\s_1}{2}+A^3\frac{\s_3}{2})\psi _1-\frac 14\, F^{\m\n
      a}F_{\m\n}^a\Big) \ ,\\ && \ncr
      S_1&=&-\, \frac {1}{16}\, \theta^{\r\s}\Delta^{\m\n\a}_{\r\s\b}\int
      d^4x\,\Big(i\bar\psi_1\g^\b\big(
      (F_{\m\n}^1\frac{\s_1}{2}+F_{\m\n}^3\frac{\s_3}{2})
      \overrightarrow{\pa}_\a-\overleftarrow{\pa}_\a
      (F_{\m\n}^1\frac{\s_1}{2}+F_{\m\n}^3\frac{\s_3}{2})\big)\psi_1\ncr
      &+& i\bar\psi_2\g^\b\big(
      (F_{\m\n}^1\frac{\s_1}{2}+F_{\m\n}^3\frac{\s_1}{2})
      \overrightarrow{\pa}_\a-\overleftarrow{\pa}_\a
      (F_{\m\n}^1\frac{\s_1}{2}+F_{\m\n}^3\frac{\s_3}{2})\big)\psi_2\ncr
      &-&
      \bar\psi_1\g^\b\frac{\s_2}{2}(F_{\m\n}^2\overrightarrow{\pa}_\a-
      \overleftarrow{\pa}_\a F_{\m\n}^2)\psi_2 +
      \bar\psi_2\g^\b\frac{\s_2}{2}
      (F_{\m\n}^2\overrightarrow{\pa}_\a-\overleftarrow{\pa}_\a
      F_{\m\n}^2)\psi_1\ncr &+&\frac{i}{2}\, F_{\m\n}^aA_\a ^a
      (\bar\psi_1\g^\b\psi_2-\bar\psi_2\g^\b\psi_1)\Big)\
      .\label{act2}\eea This will be the initial point for the
      quantization.

      \section{One-loop effective action}

      Background field method is one of the standard methods
      to obtain divergent and finite quantum
      contributions to the classical action \cite {peskin}. In the first
      step, one expands fields around their classical configuration,
      i.e. splits the fields into the background (classical)
       part and the quantum correction:
      \be \label{split} A_\m\to A_\m +\ca_\m\ \ ,\ \ \psi_{1,2} \to
      \psi_{1,2} +\Psi_{1,2} \ .\ee Quantum fields are denoted here by
      $\ca_\m$, $\Psi_{1,2}$. The functional integration over the
      quantum fields in the generating functional is then performed; the
      effective action, $\Gamma$, is the Legendre transformation of the
      generating functional. In the saddle-point approximation, the
      integration gives:
      \be \label{G1} \G [A_\m ,
      \psi_{1},\psi_2 ] =S[A_\m , \psi_{1},\psi_2 ]+\frac{i}{2}\,{\rm
      Sdet} \,\log S^{(2) }[A_\m ,\psi_{1},\psi_2 ]\ . \ee \noindent
      ${\rm Sdet}$ denotes the functional superdeterminant and
      $S^{(2)}$ is the second functional derivative of the classical
      action. For polynomial interactions, the second  derivative can be
      obtained from the quadratic part of the action; it is an
      expression of the type: \be \int d^4x \, {\pmatrix { \ca_\m^f &
      \bar \Psi_1 & \bar \Psi_2 \cr} }\,\cb\, {\pmatrix{ \ca_\n^e \cr
      \Psi_1 \cr \Psi_2 \cr}}\ ,\label{lf} \ee where we wrote $\cb$
      instead of $S^{(2)}$ as, in fact, we include the gauge fixing
      term, too. In our case the gauge fixing term is \be\label{GF}
      S_{\rm{GF}}=-\frac 12\int d^4x(D_\m\ca^{\m a})^2\ ,\ee  $D_\m$ is
      the background covariant derivative, $D_\m\ca^{\n\a}=\pa_\m\ca
      ^{\n a}+\e^{abc}A_\m^b\ca^{\n c}\ .$
      To calculate the
      one-loop correction  $$\G^{(1)} =\frac {i}{2}\,\log\Sdet \cb
      = \frac {i}{2}\,\STr\log\cb $$ perturbatively, one usually
      expands
      $\log\cb$.
      In correspondence with the notation  for the fields,
       $\cb$ is a
      3$\times$3 block matrix  $$\cb =\pmatrix{ \cb _{11} &\cb _{12}
      &\cb _{13} \cr \cb _{21} &\cb _{22} &\cb _{23} \cr \cb _{31} &\cb
      _{32} &\cb _{33} \cr } \ .$$ The submatrices $\cb _{12}$, $\cb
      _{13}$, $\cb _{21}$ and $\cb _{31}$ are Grassmann-odd while the
      rest are Grassmann-even; the supertrace is defined by $ \STr\cb =
      \Tr\cb _{11}-\Tr\cb _{22} -\Tr\cb _{33}$.  $\cb$ depends  on the
      classical fields. One should  keep in mind that $A_\m^a$ (for
      $a=1,2,3$) is a triplet, while $\psi _1$  and $\psi_2$ are dublets
      of the $SU(2)$ group.

In the absence of the interaction, $\cb$ is the inverse
propagator; in  general case, one can separate the
     kinetic part:
      $$\cb =\pmatrix{ \frac 12g_{\a\b}\d ^{ab}\Box &0 &0 \cr 0 &i\slash
      \pa &0 \cr 0&0 &i\slash \pa \cr }+\cm \ .$$
      We are to expand $\log \cb$ around identity
      $\ci ={\rm diag}
      (g_{\m\n}\d^{ab},1,1)$. To achieve this, we multiply $\cb$  by
      $\cc$ \cite{bv}: $$\cc =\pmatrix{ 2 & 0 & 0 \cr 0 & -i\slash\pa &
      0  \cr 0 & 0 & -i\slash\pa \cr} \ ,$$ and then for the one-loop
      correction we obtain   \bea\G^{(1)} &=&\frac {i}{2}\,\STr\log (\cb
      \cc)+\frac {i}{2}\,\STr\log \cc^{-1}\ncr &=&\frac {i}{2}\,\STr\log
      ( \ci +\Box ^{-1}\cm\cc )+ \frac {i}{2}\,\STr\log \cc^{-1}+\frac
      {i}{2}\,\STr\log \Box \ .\label{oloopc}\eea
      The second and the third terms, being independent on the fields,
      can be included in the
      infinite
      renormalization. Note that now the propagator for all fields is
      $\Box ^{-1}$. The operator $\Box ^{-1}\cm\cc$  defines the rules
      in the perturbation expansion.

      To get the structure of the expansion more clearly, we decompose
      $\cm\cc$ into the sum \be
      \label{mc}\cm\cc=N_0+N_1+N_2+T_1+T_2+T_3\ \ee with respect to the
      number of fields --  indices denote their number in a given term.
      $N_0$, $N_1$ and $N_2$ originate from the commutative theory,
      while $T_1$, $T_2$ and $T_3$ are the noncommutative interactions
      linear in $\theta$. One can read $\, N_0\dots T_3\, $ from
      the action (\ref{act1}-\ref{act2}), after the separation of the
      part quadratic in the quantum fields.

      In the $\theta^0$ order  we get
      \be \label{n0} N_0=\pmatrix {0&
      0&0\cr 0&i m \slash\pa & 0\cr 0 &0 & i m\slash\pa }\ .\ee

      $N_1$ is lengthy. If we write it as
      \be N_1=\pmatrix{N&
      u_{12} &u_{13}\cr u_{21}& u_{22}&u_{23}\cr u_{31} & -u_{23}&
      u_{22}\cr}\ ,\ee its matrix elements are given by
      \bea
      N_{\m\n}^{fe}&=&2\e^{efb}(\pa_\m A_\n^b-\pa_\n
      A_\m^b)+\e^{efb}g_{\m\n}(A^{\a
      b}\overrightarrow{\pa}_\a-\overleftarrow{\pa}_\a A^{\a b})\ncr
      u_{12}&=&\pmatrix{-\bar\psi_2\frac{\s_1}{2}&-i\bar\psi_1
      \frac{\s_2}{2}&-\bar\psi_2\frac{\s_3}{2}\cr}^T\g^\m\slash \pa \ncr
      u_{13}&=&\pmatrix{\bar\psi_1\frac{\s_1}{2}&-i\bar\psi_2
      \frac{\s_2}{2}&\bar\psi_1\frac{\s_3}{2}\cr}^T\g^\m\slash \pa\ncr
      u_{21}&=&2\g^\n \pmatrix{i\frac{\s_1}{2}\psi_2&\frac{\s_2}{2}
      \psi_1&i\frac{\s_3}{2}\psi_2\cr}\ncr
      u_{31}&=&2\g^\n
      \pmatrix{-i\frac{\s_1}{2}\psi_1&\frac{\s_2}{2}\psi_2&-i
      \frac{\s_3}{2}\psi_1\cr}\ncr
      u_{22}&=&-i\slash A^2\frac{\s _2}{2}\slash\pa \ncr
      u_{23}&=&(\slash A^1\frac{\s_1}{2}+\slash A^3\frac{\s_3}{2})
      \slash\pa \nonumber \ .\eea

      The remaining, $N_2$, is
      \be
      \label{n2}N_2=\pmatrix{M&0&0\cr0&0&0\cr0&0&0\cr}\ , \ee with
      $$ M^{fe}_{\m\n}=-g_{\m\n}\d^{\e
      f}A_\a^aA^{\a a}+g_{\m\n}A_\a^eA^{f\a }+2A_\m^eA_\n^f
      -2A_\m^fA_\n^e\ .$$

      We will introduce $T_1$, $T_2$ and $T_3$ symbolically;
      the full expressions are given in the Appendix. They are of the form \be
      T_1=\pmatrix{0& p_{12} & p_{13}\cr
      p_{21} &  p_{22} & p_{23} \cr p_{31} & -p_{23} & p_{22} \cr }\ ,\ee
      \be T_2=\pmatrix{Q&q_{12}&q_{13}\cr
      q_{21}&q_{22}&q_{23}\cr q_{31}&-q_{23}&q_{22}\cr }\, ,\ \
      T_3=\pmatrix{R&r_{12}&r_{13}\cr r_{21}&0&r_{23}\cr
      r_{31}&-r_{23}&0\cr }.\ee
 As already stressed, the representation of $SU(2)$ is not real
      (symmetric), and the interaction of the gauge fields with the
      Majorana spinors cannot be written in a manifestly covariant way. But all
      given operators have the same structure which is the consequence of
      the fact that the action was originally  in Dirac spinors. Note
      that for the massless fermions $N_0=0\,$; $N_1$, $T_1$ and $T_2$
      drastically simplify, as well.

      \section{Divergencies, 2-point functions}

      Introducing the decomposition (\ref{mc}), for the one-loop
      correction (\ref{oloopc}) we get \bea \label{strexp}
      \G^{(1)}&=&\frac {i}{2}\,\STr\log\,(\ci+\Box^{-1}\cm\cc)\\
      &=&\frac {i}{2}\, \sum _{n=1}^\infty {(-)^{n+1}\over n}\,
      \STr\big(\Box ^{-1}N_0 +\Box ^{-1}N_1 +\Box ^{-1}N_2 +\Box
      ^{-1}T_1 +\Box ^{-1}T_2+\Box ^{-1}T_3 \big) ^n \ \nonumber.\eea
      Our notation allows to extract the
      contributions from different terms in the expansion (\ref{strexp})
      easily. Parts of the effective action which give the 2-point
      functions have two classical fields. This means that the sum of
      indices in the monomials which we are interested in is equal to 2.
      Since
      there is an operator with the index 0, in principle, infinitely many
      terms contribute to the $n$-point functions.  However, as we
      are calculating the divergencies only, we will
      need just finite number of terms: $\Box ^{-1}N_0$ behaves as
      $p^{-1}$ in the momentum space, and the integrals become convergent
for $(\Box ^{-1} N_0)^k$  of a high enough degree.

      For the 2-point function in the zero-th order, power counting
      gives that the traces which contribute are: $\STr
      (\Box^{-1}N_1\Box^{-1}N_1) $,  $\STr (\Box^{-1}N_0\Box^{-1}N_2) $
      and $ \STr (\Box^{-1}N_0(\Box^{-1}N_1)^2)$. Performing the Fourier
      transformation and the dimensional regularization we obtain
      \bea \STr
      (\Box^{-1}N_1\Box^{-1}N_1)&=& \frac{i}{(4\pi)^2\e}\int d^4x
      \,\Big(-24A_\m^a\Box A^{\m a}-24(\pa_\m A^{\m
      a})^2+6i\bar\psi\slash\pa\psi\Big)\ ,\ncr \STr
      (\Box^{-1}N_0\Box^{-1}N_2)\,&=&0\ ,\ncr \STr
      (\Box^{-1}N_0(\Box^{-1}N_1)^2)&=&\frac{i}{(4\pi)^2\e} \int d^4 x\,
      12m\bar\psi\psi\ .\nonumber\eea Adding, we get the standard result
      of the commutative theory
      \bea\G_{2A,2\psi}
      &=&-\frac {i}{4}\, \STr (\Box^{-1}N_1\Box^{-1}N_1)+\frac {i}{2}\,
      \STr (\Box^{-1}N_0(\Box^{-1}N_1)^2)\ncr &=& \frac{1}{
      (4\pi)^2\e}\Big(-6A_\m^a\Box A^{\m a}-6(\pa_\m A^{\m
      a})^2+\frac{3i}{ 2}\bar\psi\slash\pa\psi-6m\bar\psi\psi\Big)\ .
      \eea The contribution of the ghost action (\ref{GF}) should be
      added, too: \be \G_{2,gh}=\frac {1}{(4\pi)^2\e}\, \frac13\,\int
      d^4x\, F_{\m\n}^a F ^{\m\n a}\ .\ee

      Now we
      consider the $\theta$-linear order. Potentially divergent terms
      are\hfill\break  $\STr(\Box^{-1}N_1\Box^{-1}T_1)$, $\STr
      ((\Box^{-1}N_0\Box^{-1}N_1+\Box^{-1}N_1\Box^{-1}N_0)\Box^{-1}T_1)
      $, $ \STr(\Box^{-1}N_0\Box^{-1}T_2) $ and
      $\STr((\Box^{-1}N_0)^2\Box^{-1}N_1+\Box^{-1}N_0\Box^{-1}N_1\Box^{-1}N_0+
      \Box^{-1}N_1(\Box^{-1}N_0)^2)\Box^{-1}T_1) $.\hfill\break The
      dimensional regularization gives the following:
      $$ \STr(\Box^{-1}N_1\Box^{-1}T_1)=\frac{i}{
      (4\pi)^2\e}\theta^{\m\n}\Big( \frac{i}{ 4}\,
      \e_{\m\n\r\s}\bar\psi\g_5\g^\s\Box\pa^\r\psi+ \frac{1}{ 2}\,
      m\bar\psi\s_{\n\a}\pa_\m\pa^\a\psi-\frac{1}{ 4}\,
      m\bar\psi\s_{\m\n}\Box\psi\Big)\ ,$$ $$ \STr
      ((\Box^{-1}N_0\Box^{-1}N_1+\Box^{-1}N_1\Box^{-1}N_0)\Box^{-1}T_1)=0\
      ,$$  $$ \STr(\Box^{-1}N_0\Box^{-1}T_2)=0\ ,$$ and
      \bea&&\STr\Big(((\Box^{-1}N_0)^2\Box^{-1}N_1+\Box^{-1}N_0\Box^{-1}N_1\Box^{-1}N_0+
      \Box^{-1}N_1(\Box^{-1}N_0)^2)\Box^{-1}T_1\Big)\ncr && =\frac{i}{
      (4\pi)^2\e}\,\frac 34\, \theta^{\m\n}\Big(-im^2
      \e_{\m\n\a\b}\bar\psi\g_5\g^\b\pa^\a\psi +
      m^3\bar\psi\psi\s_{\n\m}\psi\Big)\ .\nonumber\eea

      All together, the divergent part of the 2-point function in the
      linear order reads: \bea\G_{2\psi}^\prime &=&\frac{1}{
      (4\pi)^2\e}\,\frac 12\,\theta^{\m\n}\Big(\frac{i}{ 4}\,
      \e_{\m\n\r\s}\bar\psi\g_5\g^\s\Box\pa^\r\psi+\frac{1}{ 2}\,
      m\bar\psi\s_{\n\a}\pa_\m\pa^\a\psi-\frac{1}{ 4}\,
      m\bar\psi\s_{\m\n}\Box\psi\Big)\ncr &+&\frac{1}{
      (4\pi)^2\e}\,\frac{3}{8 }\, \theta^{\m\n}\Big(-im^2
      \e_{\m\n\a\b}\bar\psi\g_5\g^\b\pa^\a\psi
      +m^3\bar\psi\s_{\n\m}\psi\Big) .\label{2psi}\eea

      This is a nice result: comparing (\ref{2psi}) with the
      $\theta$-linear correction for the 2-point functions in NC QED
      \cite{mv}, we see that in the $SU(2)$ case also, only  fermionic
      propagator gets a correction; the correction is, up to a factor,
      the same as  the one for
      $U(1)$. This has further consequences. In NC QED we argued that
      the massive terms obstruct renormalization, as only for the case
      $m=0$ one can redefine fields in such a way that the divergent
      terms disappear. The  analysis can be repeated for $SU(2)$ without
      change. Hence, we come to the conclusion: the NC gauge theories with
      massive fermions are not renormalizable.

      For this reason in the calculations of 3-point and 4-point
      functions we focus to the massless case. One might add that the
      calculations become so cumbersome that otherwise
      they would hardly be doable.

      \section{3-point and 4-point functions}

      The background field method is a gauge covariant method and
      therefore it gives the covariant results. On the other hand, the
      separation into 2-point, 3-point etc. functions breaks the gauge
      covariance: for instance, $\G_{2\psi}^\prime $ given by the
      formula (\ref{2psi}) is not covariant; it is  a part of the
      covariant expression (written assuming that $m=0$) which is,
       up to the order of the covariant derivatives, equal to: \be
      \G_2^\prime = \frac{1}{ (4\pi)^2\e}\,\frac{i}{
      8}\,\theta^{\m\n}\e_{\m\n\r\s}\bar\psi\g_5\g^\s D^2D^\r\psi
      ,\label{2gama}\ee
 Writing  $   \G_2^\prime$ in the form (\ref{2gama}), we included the
 parts of the 3-point,  4-point and 5-point
      functions. When one calculates the 3-point functions,
      from the dimensional analysis it is easy to see that, apart from the terms
      residual from (\ref{2gama}), basically only two terms
      contribute (up to the partial integration and  various
      combinations of indices). They are the leading terms in the
      covariant expressions \be\label{form}\theta \bar\psi \g F(D\psi )\
      \ {\rm and}\ \ \ \theta\bar\psi \g (DF)\psi \ ,\ee ($\g$ stands
      for the  products of the $\g$-matrices). However, as we stressed
      already, the calculation i.e. the organization of terms becomes
      increasingly difficult, so in order to find the 3-point functions
      we use a trick. We calculate the coefficients of the
      terms (\ref{form}) in the 4-point functions. When we are doing this, we
      can assume that the background spinor field is constant, so the
      covariant derivative $D_\m\psi$ reduces to $A_\m \psi$; in this case
            $\, N_1\dots T_3\,$ also simplify. The gauge covariance enables us to
      recover the result for  the 3-point functions  uniquely at the end of the calculation.

      Divergent parts of the
      3-point functions are in the terms
      $\STr((\Box^{-1}N_1)^2\Box^{-1}T_1)$ and $
      \STr(\Box^{-1}N_1\Box^{-1}T_2) $; the corresponding traces
in the 4-point functions are \hfill\break
      $\STr(\Box^{-1}N_1\Box^{-1}T_3)$ and $
      \STr((\Box^{-1}N_1)^2\Box^{-1}T_2)$ . The
divergent part of the
      first trace is
      \be\STr(\Box^{-1}N_1\Box^{-1}T_3)=-\frac{i}{ (4\pi)^2\e}\,
      \theta^{\m\n}\Big( 6A_\n^a F^a_{\m\a}\bar\psi\g^\a\psi- 3A_\a^a
      F^a_{\m\n}\bar\psi\g^\a\psi\Big)\ ,\ee while for the second one we
      get \bea\STr((\Box^{-1}N_1)^2\Box^{-1}T_2)&=&\frac{i}{
      (4\pi)^2\e}\,\theta^{\m\n}\Big(\frac{33}{ 24}\,A_\m^a (\pa_\n
      A_\a^a-\pa_\a A_\n^a)\bar\psi\g^\a\psi\ncr &-&\frac{5}{
      16}\,A_\a^a (\pa_\n A_\m^a-\pa_\m A_\n^a)\bar\psi\g^\a\psi+\frac
      12A^{a\a} (\pa_\n A_\a^a-\pa_\a A_\n^a)\bar\psi\g_\m\psi\ncr &
      +&\e^{abc}\big(-\frac {5}{6}\,\e_{\n\a\b\r}A^{a\a}(\pa_\m
      A^{b\b} -\pa^\b A_\m^b)\bar\psi
      \frac{\s_c}{2}\,\g_5\g^\r\psi\ncr&-&\frac{25}{ 24}\,\e_{\n\a\b\r}
      A_\m^a (\pa^\a A^{c\b}-\pa^\b A^{c\a})\bar\psi
      \frac{\s_b}{2}\,\g_5\g^\r\psi\ncr&-&\frac {1}{16}\,\e_{\n\m\a\b}A_\s^c
      (\pa^\b A^{a\a} -\pa^\a A^{a\b})\bar\psi
      \frac{\s_b}{2}\,\g_5\g^\s\psi\ncr &-&\frac{1}{
      8}\,\e_{\n\m\a\b}A^{b\a}
\pa_\s A^{a\s }\bar\psi \frac{\s_c}{2}\,\g_5\g^\b\psi\big)\Big)\ .
      \eea

      When we add these expressions and try to 'covariantize' the
      result, we obtain a piece which does not match: it is precisely
      the part of  the 2-point function (\ref{2gama}). The rest gives
      the 3-point function in its covariant form:
      \bea\label{3gama}\G^{\prime}_3&=&-\frac{1}{
      (4\pi)^2\e}\,\frac 12\,\theta^{\m\n}\Big(-\frac{111i}{
      6}\,\bar\psi\g^\a F_{\n\a}D_\m\psi\ncr &-&\frac{43i}{4}\,\bar\psi\g^\a
      F_{\m\n}D_\a\psi-\frac{3i}{4}\,\bar\psi\g^\a (D_\a F_{\m\n})\psi\ncr &+&
       2i\bar\psi\g_\m F_{\n\a}D^\a\psi+i\bar\psi\g_\m (D^\a
      F_{\n\a})\psi\ncr &+&\frac{5}{ 8}\,\e_{\n\a\b\r}
      \bar\psi\g_5\g^\r (D_\m F^{\a\b})\psi-\frac
      {1}{16}\,\e_{\m\n\a\b}\bar\psi\g_5\g^\s (D_\s F^{\a\b})\psi\ncr
      &+&\frac {1}{8}\,\e_{\m\n\a\b}\big(2\bar\psi\g_5\g^\b
      F^{\r\a}D_\r\psi+\bar\psi\g_5\g^\b(D_\r F^{\r\a})\psi\big)\Big)\ .
      \eea

4-fermionic vertex
      has a very important role in the discussion of
      renormalizability.
 The corresponding 4-point function is relatively easy
      to find: to this end one can put $A_\m^a = 0$ in $\, N_1\dots
      T_3\,$. The
      divergent part comes from $\STr ((\Box^{-1}N_1)^2\Box^{-1}T_2)$
      and  $\STr ((\Box^{-1}N_1)^3\Box^{-1}T_1)$; the final result is
      \be \label{4gama}\G^\prime_{4\psi} =
      \frac{1}{
      (4\pi)^2\e}\,\frac{9}{32}\,\theta^{\m\n}\e_{\m\n\r\s}\bar\psi\g_5\g^\s\psi
      \, \bar\psi\g^\r\psi \ .\ee

      (\ref{2gama}), (\ref{3gama}) and (\ref{4gama}) are the main
      results of our calculation.

      \section{Discussion}

As we mentioned in the introduction, there is no a priori
criterion which would fix the nonuniqueness in the SW map. The
redefinition of fields allowed by it changes the action; the terms
which appear are of the forms:
 \be \label{DSA} \Delta
      S_{A}^{(n)} =\int d^4x\, (D_\n F^{\m\n}){\bf A}_\m^{(n)}\ ,\ee \be
      \label{DSpsi} \Delta S_{\psi}^{(n)} =\int d^4x\, \Big( \bar\psi
      i\slash D {\bf \Psi}^{(n)}+ \bar {\bf \Psi}^{(n)}i\slash D\psi
      \Big)\ ,\ee   written for the case of massless fermions.  ${\bf
      A}_\m^{(n)}$ and ${\bf \Psi} ^{(n)}$ are gauge covariant expressions of
     the $n$-th order in $\theta$. The important thing in
      (\ref{DSA}-\ref{DSpsi}) is that, besides $F^{\m\n}$ and $\psi$,
      they  contain at least one derivative.

The divergencies (\ref{2gama}), (\ref{3gama}), (\ref{4gama}) which
we obtained are such that they cannot be subtracted by the usual
counterterms. However, if they were of the types
(\ref{DSA}-\ref{DSpsi}), one could  include them in the field
redefinitions; thus the theory would be renormalizable in a
generalized sense.
   Analyzing the divergencies, we see that the
       situation with the NC $SU(2)$ is
      pretty much the same as with the  electrodynamics.
       We  already noted that
      the propagator correction (\ref{2gama}) breaks the
      renormalizability, unless $m=0$.  The 3-point functions
      in the massless case present no
      problem, too. Gluon  3- and 4-vertices get no quantum
      correction in the $\theta$-linear order --  there is no classical
      gluon vertex in (\ref{act2}) in that order, either. In this
      respect the behavior is again similar to  $U(1)$, where
      $\theta$-linear 3-photon vertices did exist: quantum
      one-loop corrections were precisely of the same form as the
     corresponding classical vertices \cite{w}. Further, the
fermion-gluon 3-vertex (\ref{3gama}) is already written in the
form  (\ref{DSpsi}) with
\be{\bf \Psi}
^{(1)}=\theta^{\m\n}\Big(\kappa_1F_{\m\n}+i\kappa_2\s_{\m\r}F_\n{}^\r
+i\kappa_3\e_{\m\n\r\s}\g_5F^{\r\s}+\kappa_4\s_{\m\n}D^2\Big)\psi
\ .\ee We observe that the
         fermions in the renormalized theory would be redefined
 via the gluon fields, i.e.,  noncommutativity  would be mixed with
 (or partly immersed into) the gauge interactions.

    However, there is a divergent term
      which spoils the renormalizability: the 4-point function (\ref{4gama}).
      It predicts the
      current-current interaction, and there is no simple way to
      circumvent this coupling induced by noncommutativity.

      In fact, from the dimensional analysis we can see that in the
      massless case, propagators in NC gauge theories are renormalizable
      to all orders. Namely, for gauge bosons the $n$-th order
      corrections are of the form \be \underbrace{\theta\dots\theta}_n\,
      A\, \underbrace{\pa\dots\pa}_k\, A\ ,\label{AA}\ee while for
      fermions they are
      \be \underbrace{\theta\dots\theta}_n\,
      \bar\psi\gamma\, \underbrace{\pa\dots\pa}_k\, \psi\
      .\label{psipsi}\ee The power counting gives  the number of
      derivatives $k$: for the gluon propagator  $k=2+2n$; for
       the fermions, $k=1+2n$. This shows that one can, in all orders, transform
      (\ref{AA}-\ref{psipsi})  into the desired forms
      (\ref{DSA}-\ref{DSpsi}). The use of the
      background field method  guarantees the
      gauge covariance.

      On the other hand, the vertices are potentially problematic. From
the  power counting we see that in the linear order the 'wrong'
      vertex could be
      \be\theta(\bar\psi\gamma\psi)^2\label{theta1}\ ,\ee while in the
      quadratic order we could have, e.g., \be\theta^2F^4\ ,
      \ \  \theta^2(\bar\psi\gamma\psi)^2F\ .\label{theta2}\ee These terms contain
      no derivatives and therefore break the SW generalized
renormalization scheme.
      The term
      (\ref{theta1}) is present for both $U(1)$ and $SU(2)$ theories
       coupled to fermions. An
      interesting fact, however, is that in both theories it has the
      same form
      (a different coefficient), namely \be
      \theta^{\m\n}\e_{\m\n\r\s}\bar\psi\g_5\g^\s\psi \,
      \bar\psi\g^\r\psi \ ,\ee whereas the other combinations allowed by
      covariance, as, e.g.
      $\theta^{\m\n}\e_{\m\n\r\s}\bar\psi\g_5\g^\s\frac{\s_a}{2}\,\psi
      \, \bar\psi\g^\r\frac{\s_a}{2}\psi$ or
      $\theta^{\m\n}\bar\psi\g_5\g_\m\psi \, \bar\psi\g_\n\psi$,
      never show up. This opens the possibility that this
      divergence
      might cancel in a gauge theory based on the product of
      gauge groups.

However, we are not  in  favor of a theory which needs too much
fine
      tuning. Thus we are inclined to interpret our results (and the previous ones,
      \cite{u1,w,mv}) as an indication that  NC  gauge theories
      coupled to fermions are not renormalizable. But before
a definite conclusion, one should certainly check
      whether the specific $\theta^2$-corrections, as $4A$
      vertex (\ref{theta2}),
      vanish. The presence of the $\theta ^2F^4$
      divergence would  prove non-renormalizability, possibly even for
      the pure gauge theories. It would also be interesting to understand if
      there is some
      systematics in
      the behavior of various divergent terms.

      \section{Appendix}

      We present here the operators from the expansion
  (\ref{mc}) which are induced by the $\theta$-linear
      interaction terms. The matrix $T_1$  containing one background
      field is
       \be\label{t1} T_1=\pmatrix{0& p_{12} &
      p_{13}\cr
      p_{21} &  p_{22} & p_{23} \cr p_{31} & -p_{23} & p_{22} \cr }\ ,\ee
      where the matrices $p_{ij}$ are given by
      \bea p_{12}&=&-\frac {i}{4}\,\theta^{\r\s}\pmatrix{-m
      \d^\n_\s\pa_\r\bar\psi_1\s _1+\frac
      {i}{2}\,\D^{\m\n\a}_{\r\s\b}(\pa_\m\bar\psi_1)
      \s_1\g^\b\pa_\a&\cr i\d^\n_\s m\pa_\r\bar\psi_2\s _2+{1\over
      2}\D^{\m\n\a}_{\r\s\b}(\pa_\m\bar\psi_2)
      \s _2\g^\b\pa_\a\cr
      -m\d^\n_\s\pa_\r\bar\psi_1\s
      _3+\frac{i}{2}\,\D^{\m\n\a}_{\r\s\b}(\pa_\m\bar\psi_1) \s
      _3\g^\b\pa_\a\cr}\slash \pa \ , \ncr &&\ncr
      p_{13}&=&-\frac{i}{4}\,\theta^{\r\s}\pmatrix{-m\d^\n_\s
      \pa_\r\bar\psi_2\s _1+\frac {i}{2}\,\D^{\m\n\a}_{\r\s\b}(\pa_\m\bar\psi_2)
      \s _1\g^\b\pa_\a&\cr
      -im\d^\n_\s\pa_\r\bar\psi_1\s
      _2-\frac{1}{2}\D^{\m\n\a}_{\r\s\b}(\pa_\m\bar\psi_1)
      \s _2\g^\b\pa_\a\cr
      -m\d^\n_\s\pa_\r\bar\psi_2\s _3+\frac{i}{
      2}\,\D^{\m\n\a}_{\r\s\b}(\pa_\m\bar\psi_2) \s
      _3\g^\b\pa_\a\cr}\slash \pa \ ,\ncr && \ncr
      p_{21}&=&\frac{1}{2}\,\theta^{\r\s}\pmatrix{m\d^\n_\s\s_1\psi_1\pa_\r-
      \frac{i}{2}\D^{\m\n\a}_{\r\s\b}\g^\b \s_1(\pa_a\psi_1)\pa_\m\cr
      im\d^\n_\s\s
      _2\psi_2\pa_\r+\frac 12\D^{\m\n\a}_{\r\s\b}\g^\b
      \s_2(\pa_a\psi_2)\pa_\m\cr
      m\d^\n_\s\s_3\psi_1\pa_\r-\frac{i}{2}\,\D^{\m\n\a}_{\r\s\b}\g^\b
      \s_3(\pa_a\psi_1)\pa_\m\cr}^T\ncr &&\ncr
      p_{31}&=&\frac12\,\theta^{\r\s}\pmatrix{m\d^\n_\s\s_1\psi_2\pa_\r-\frac{i}{2}\,\D^{\m\n\a}_{\r\s\b}\g^\b
      \s_1(\pa_a\psi_2)\pa_\m\cr -im\d^\n_\s\s_2\psi_1\pa_\r-\frac
      12\,\D^{\m\n\a}_{\r\s\b}\g^\b \s_2(\pa_a\psi_1)\pa_\m\cr
      m\d^\n_\s\s_3\psi_2\pa_\r-\frac{i}{ 2}\,\D^{\m\n\a}_{\r\s\b}\g^\b
      \s_3(\pa_a\psi_2)\pa_\m\cr}^T\ncr
      p_{22}&=&-\frac {i}{4}\,\theta^{\r\s}\Big(m((\pa_\r
      A_\s^1)\s_1+(\pa_\r A_\s^3)\s_3)
      -\frac{i}{2}\,\D^{\m\n\a}_{\r\s\b}\g^\b((\pa_\m
      A_\n^1)\s_1+(\pa_\m A_\n^3)\s_3)\pa_\a \Big)\slash \pa \ncr
      p_{23}&=&\frac{1}{4}\,\theta^{\r\s}\Big(-m(\pa_\r A_\s^2)\s_2-\frac
      {i}{2}\,
      \D^{\m\n\a}_{\r\s\b}\g^\b(\pa_\m A_\n^2)\s_2\pa_\a\Big)\slash\pa\
      .\eea

      The matrix $T_2$ is
      \be T_2=\pmatrix{Q&q_{12}&q_{13}\cr
      q_{21}&q_{22}&q_{23}\cr q_{31}&-q_{23}&q_{22}\cr }\ee with $\
      Q=\pmatrix{a&b&c\cr -b&a&d\cr-c&-d&a\cr }\ ,\ $ and
      \bea a &=&\frac{1}{
      16}\,\theta^{\r\s}\Delta^{\m\n\a}_{\r\s\b}\Big(\pa_\a(\bar\psi\g^\b\psi)+\bar\psi
      \g^\b\psi\pa_\a\Big)\ncr  b&=&\frac{m}{
      4}\,\theta^{\m\n}(\bar\psi_1\s_3\psi_1+\bar\psi_2\s_3\psi_2)+\frac{i}{
      8}\,\theta^{\r\s}\Delta^{\m\n\a}_{\r\s\b}\Big((\pa_\a\bar
      \psi_1)\g^\b \s_3\psi_1+(\pa_\a\bar \psi_2)\g^\b
      \s_3\psi_2\Big)\ncr  c&=&-\frac{im}{
      4}\,\theta^{\m\n}(\bar\psi_1\s_2\psi_1-\bar\psi_2\s_2\psi_2)-\frac{1}{
      8}\,\theta^{\r\s}\Delta^{\m\n\a}_{\r\s\b}\Big((\pa_\a\bar
      \psi_1)\g^\b \s_2\psi_1+(\pa_\a\bar \psi_2)\g^\b
      \s_2\psi_2\Big)\ncr  d&=&\frac{m}{
      4}\,\theta^{\m\n}(\bar\psi_1\s_1\psi_1+\bar\psi_2\s_1\psi_2)+\frac{i}{
      8}\,\theta^{\r\s}\Delta^{\m\n\a}_{\r\s\b}\Big((\pa_\a\bar
      \psi_1)\g^\b \s_1\psi_1+(\pa_\a\bar \psi_2)\g^\b \s_1\psi_2\Big)\
      , \ \nonumber \eea while the other matrix elements are \bea
      q_{12}&=&-\frac{i}{
      8}\,\theta^{\r\s}\pmatrix{-m\d^\n_\s(\bar\psi_1A_\r^2\s_3+i\bar\psi_2A_\r^3\s_2)
      -\frac {i}{2}\,\Delta^{\m\n\a}_{\r\s\b}((\pa_\m
      A_\a^1)\bar\psi_2+\pa_\m A_\a^1\bar\psi_2)\g^\b\cr -i
      m\d^\n_\s(-i\bar\psi_1A_\r^3\s_1+i\bar\psi_1A_\r^1\s_3)-\frac{i}{
      2}\,\Delta^{\m\n\a}_{\r\s\b}((\pa_\m A_\a^2)\bar\psi_2+\pa_\m
      A_\a^2\bar\psi_2)\g^\b\cr
      -m\d^\n_\s(-\bar\psi_1A_\r^2\s_1-i\bar\psi_2A_\r^1\s_2)-\frac{i}{
      2}\,\Delta^{\m\n\a}_{\r\s\b}((\pa_\m A_\a^3)\bar\psi_2+\pa_\m
      A_\a^3\bar\psi_2)\g^\b\cr}\slash\pa \ncr &&\ncr &-&\frac{i}{
      16}\,\theta^{\r\s}
      \Delta^{\m\n\a}_{\r\s\b}\pmatrix{-iA_\m^2((\pa_\a\bar\psi_1)-\bar\psi_1\pa_\a)\s_3)
      +A_\m^3((\pa_\a\bar\psi_2)-\bar\psi_2\pa_\a)\s_2)\cr
      iA_\m^3((\pa_\a\bar\psi_1)-\bar\psi_1\pa_\a)\s_1
      +iA_\m^1((\pa_\a\bar\psi_1)-\bar\psi_1\pa_\a)\s_3) \cr
      iA_\m^2((\pa_\a\bar\psi_1)-\bar\psi_1\pa_\a)\s_1
      -A_\m^1((\pa_\a\bar\psi_2)-\bar\psi_2\pa_\a)\s_2)
      \cr}\g^\b\slash\pa \ncr &&\ncr
      q_{13}&=&q_{12}|_{\bar\psi_1\to\bar\psi_2,\
      \bar\psi_2\to-\bar\psi_1}\ncr & &\ncr
       q_{21}&=&\frac {1}{4}\,\theta^{\r\s}
       \pmatrix{m\d^\n_\s(-A_\r^2\s_3\psi_1+iA_\r^3\s_2\psi_2)+
       \frac{i}{2}\,\Delta^{\m\n\a}_{\r\s\b}\g^\b\psi_2((\pa_\m
      A_\a^1)-A_\a^1\pa_\m)\cr
      m\d^\n_\s(-A_\r^3\s_1\psi_1+A_\r^1\s_3\psi_2)+\frac{i}{2}\,
      \Delta^{\m\n\a}_{\r\s\b}\g^\b\psi_2((\pa_\m
      A_\a^2)-A_\a^2\pa_\m)\cr
      m\d^\n_\s(-A_\r^2\s_1\psi_1-iA_\r^1\s_2\psi_2)+\frac{i}{
      2}\,\Delta^{\m\n\a}_{\r\s\b}\g^\b\psi_2((\pa_\m
      A_\a^3)-A_\a^3\pa_\m)\cr }^T \ncr &&\ncr
      &+& \frac {1}{8}\,
      \theta^{\r\s}\Delta^{\m\n\a}_{\r\s\b}\g^\b\pmatrix{i\s_3
      (\pa_\a\psi_1+(\pa_\a\psi_1))A_\m^2+\s_2
      (\pa_\a\psi_2+(\pa_\a\psi_2))A_\m^3\cr
      i\s_1(\pa_\a\psi_1+(\pa_\a\psi_1))A_\m^3-i\s_3
      (\pa_\a\psi_1+(\pa_\a\psi_1))A_\m^1\cr
      -i\s_1(\pa_\a\psi_1+(\pa_\a\psi_1))A_\m^2-\s_2
      (\pa_\a\psi_2+(\pa_\a\psi_2))A_\m^1\cr}^T \ncr &&\ncr
      q_{23}&=&q_{21}|_{\psi_1\to\psi_2,\ \psi_2\to-\psi_1}\ncr &&\ncr
      q_{22}&=&-i\frac{m}{4}\,\theta^{\r\s}(A_\r^2 A_\s^3\s_1+A_\r^1
      A_\s^2\s_3)\slash \pa \ncr &-&\frac{1}{
      16}\,\theta^{\r\s}\Delta^{\m\n\a}_{\r\s\b}\g^\b\Big[\s_1(\pa_\a
      A_\m^2A_\n^3+A_\m^2A_\n^3\pa_\a)+\s_3(\pa_\a
      A_\m^1A_\n^2+A_\m^1A_\n^2\pa_\a)\Big]\slash \pa\ncr &&\ncr
      q_{23}&=& \frac{m}{4}\,\theta^{\r\s}A_\r^3 A_\s^1\s_2\slash \pa
      -\frac {i}{16}\,\theta^{\r\s}\Delta^{\m\n\a}_{\r\s\b}\g^\b
      \s_2(\pa_\a A_\m^3A_\n^1+A_\m^3A_\n^1\pa_\a)\slash\pa \ncr
      &-&\frac {1}{16}\,\theta^{\r\s}\Delta^{\m\n\a}_{\r\s\b}(\pa_\m
      A_\n^a)A_\b^a\g^\b\slash \pa\ .\nonumber \eea

      Finally, the operator $T_3$ containing three fields is
      \be
      T_3=\pmatrix{R&r_{12}&r_{13}\cr r_{21}&0&r_{23}\cr
      r_{31}&-r_{23}&0\cr }\ ,\ee with

      \bea R^{fe,\m\n}&=&-\frac{3}{
      16}\,
      \theta^{\r\s}\Delta^{\m\n\a}_{\r\s\b}\e^{afe}A^a_\a\bar\psi\g^\b\psi
      \ncr
      r_{12}^a&=&-\frac{3}{32}\,\theta^{\r\s}
      \Delta^{\m\n\a}_{\r\s\b}\e^{abc}A_\m^bA_\a^c\bar\psi_2\g^\b\slash\pa\ncr
      r_{13}^a&=&\frac{3}{
      32}\,\theta^{\r\s}\Delta^{\m\n\a}_{\r\s\b}\e^{abc}A_\m^bA_\a^c\bar\psi_1\g^\b\slash
      \pa \ncr r_{21}^a&=&\frac{3i}{
      16}\theta^{\r\s}\Delta^{\m\n\a}_{\r\s\b}\e^{abc}\g^\b\psi_2A_\m^bA_\a^c\ncr
       r_{31}^a&=&-\frac{3i}{
      16}\,\theta^{\r\s}\Delta^{\m\n\a}_{\r\s\b}\e^{abc}\g^\b\psi_1A_\m^bA_\a^c\ncr
      r_{23}&=&-\frac{3}{
      16}\,\theta^{\r\s}\Delta^{\m\n\a}_{\r\s\b}A_\m^1A_\n^2A_\a^3\g^\b\slash
      \pa \ .\nonumber\eea
\vskip1cm
      {\bf Acknowledgment}\ \ This work was partly done during our visits to MPI f\"ur Physik,
      Munich and AEI f\" ur Gravitationsphysik, Potsdam; we thank Prof.
      J. Wess and Prof. H. Nicolai for their kind hospitality. Our
      research is funded by the project 1468 of the Serbian Ministry of
      Science, Technology and Development.


\begin{thebibliography}{99}

      \bibitem{phi4} S.~Minwalla, M.~Van Raamsdonk and N.~Seiberg,
      JHEP {\bf 0002} (2000) 020; I.~Y.~Aref'eva, D.~M.~Belov and
      A.~S.~Koshelev, Phys.\ Lett.\ B {\bf 476}, 431 (2000);
      I.~Chepelev and R.~Roiban, JHEP {\bf 0005} (2000) 037,
       JHEP {\bf 0103} (2001) 001;

      \bibitem{gp} L.~Griguolo and M.~Pietroni, JHEP {\bf 0105} (2001)
      032

      \bibitem{becchi} C.~Becchi, S.~Giusto and C.~Imbimbo,
      Nucl.\ Phys.\ B {\bf 633} (2002) 250;
       "The
      Renormalization of Non-Commutative Field Theories in the Limit of
      Large Non-Commutativity", preprint hep-th/0304159

      \bibitem{gw} H.~Grosse and R.~Wulkenhaar,  "Power-counting theorem
      for non-local matrix models and renormalisation", preprint
      hep-th/0305066;
       "Renormalisation of
      $\phi^4$-theory on noncommutative $\mathbb{R}^2$ in the matrix
      base", preprint hep-th/0307017

      \bibitem{U1} M.~Hayakawa,  "Perturbative analysis on infrared and
      ultraviolet aspects on noncommutative QED on $R^4$", preprint
      hep-th/9912167; Phys. Lett. B {\bf 478} (2000) 394; C.~P.~Martin and
      D.~S\'anchez-Ruiz, Phys. Rev. Lett. {\bf 83} (1999) 476; I.~F.~Riad and
      M.~M.~Sheikh-Jabbari, JHEP {\bf 0008} (2000) 045

      \bibitem{UN}  L.~Bonora, M. Schnabl, M.~M.~Sheikh-Jabbari and
      A.~Tomasiello,
      Nucl. Phys. B {\bf 589} (2000) 461; L.~Bonora and M.~Salizzoni, Phys.
      Lett. B {\bf 504} (2001) 80; A.~Armoni, Nucl. Phys. B {\bf 593} (2001) 229;
      M.~Chaichian, P.~Pre\v snajder, M.~M.~Sheikh-Jabbari and
      A.~Tureanu,  Phys. Lett. B {\bf 526} (2002) 132
      \bibitem{sw} N.~Seiberg and E.~Witten, JHEP {\bf 9909} (1999) 032

      \bibitem{smodel} X.~Calmet, B.~Jur\v co, P.~Schupp, J.~Wess and
       M.~Wohlgennant, Eur. Phys. J. C {\bf 23} (2002) 363;
      P.~Aschieri, B.~Jurco, P.~Schupp and J.~Wess, Nucl. Phys. B {\bf
      651}
      (2003) 45

      \bibitem{u1} A.~A.~Bichl, J.~M. Grimstrup, H.~Grosse, L.~Popp, M.~Schweeda
and R.~Wulkenhar, JHEP {\bf 0106} (2001) 013;

      \bibitem{w} R.~Wulkenhaar,  JHEP {\bf 0203} (2002) 024

      \bibitem{susy} A.~A.~Bichl, M.~Ertl, A.~Gerhold, J.~M.~Grimstrup,
      H.~Grosse, L.~Popp, V.~Putz, M.~Schweda and R.~Wulkenhaar,
      Non-commutative $U(1)$ Super-Yang-Mills Theory: Perturbative
      Self-Energy Corrections, preprint hep-th/0203141

       \bibitem{mv} M.~Buri\' c and V.~Radovanovi\' c, JHEP {\bf 0210} (2002)
      074

      \bibitem{peskin} M.~E.~Peskin and D.~V.~Schroeder, {\it An
      Introduction to Quantum Field Theory}, Addison-Wesley, 1995,
      Reading, MA

      \bibitem{mad} J.~Madore, {\it An Introduction to Noncommutative
      Differential Geometry and its Physical Applications}, Camb. Univ.
      Press, Cambridge, 1999

      \bibitem{mssw} J.~Madore, S.~Schraml, P.~Schupp and J.~Wess, Eur.
      Phys. Jour. C {\bf 16} (2000) 161

      \bibitem{jmssw} B.~Jur\v co, L. M\"oller, S.~Schraml, P.~Schupp and
      J.~Wess,  Eur. Phys. J. C {\bf 21} (2001) 383

      \bibitem{a} T.~Asakawa and I.~Kishimoto,  "Comments on Gauge
      Equivalence in Noncommutative Theory", preprint hep-th/9909139

      \bibitem{bv} A.~G.~Barvinsky and G.~A.~Vilkovisky , Phys. Rep. {\bf
      119}
      (1985) 1

      \end{thebibliography}
      \end{document}